# Gravity effects on a bio-inspired self-burrowing probe in granular soils

Bowen Wang[a]; Ningning Zhang[a, *]; Yuyan Chen[b]; Alejandro Martinez [b]; Raul Fuentes[a]

[a] *Institute of Geomechanics and Underground Technology, RWTH Aachen University, 52074 Aachen, Germany*

[b] *Department of Civil and Environmental Engineering, University of California Davis, Davis, CA 95616, United States*

* Corresponding author E-mail address: n.zhang@gut.rwth-aachen.de (N. Zhang)

**ABSTRACT**

In recent years, self-burrowing probes have been studied since they can be suitable for soil monitoring in locations with limited access such as outer space bodies and underneath existing structures. We study the performance of a self-burrowing probe under different gravity conditions, from low gravity (i.e., 1/6*g*, 1/3*g* and 1*g*) to high gravity (i.e., 5*g*, 10*g* and 15*g*), specifically in terms of penetration distance and energy consumption. Results show that the probe reaches efficient penetration in all gravity conditions and that it achieves larger penetration distances in high gravity conditions. However, the penetration efficiency, shown as unit energy per meter, is higher in low gravity. Additionally, we prove that a simple dimensional analysis provides reasonable scaling factors for first order effects in forces, velocities and energy. The findings in this study give confidence to the potential use of self-burrowing probes in campaigns of soil testing and sensor deployment in outer space or centrifuges in which the gravity conditions can differ from Earth.

# List of Notations

$C$      Stage Criteria

$D_C$      chamber diameter

$D_r$      relative density

$D_S$      original shaft diameter

$D_{SA}$      shaft anchor diameter

$D_{TA}$      tip anchor diameter

$d_c$      probe tip cone diameter

$ER_S$      shaft expansion ratio

$ER_T$      tip expansion ratio

$F_{N,z}$      neck vertical force

$F_{SA}$      shaft anchor capacity

$F_{S,bot,z}$      vertical force on shaft bottom surface

$F_{S,r}$      actual shaft radial force

$F_{S,r,inc}$      shaft radial force increment

$F_{S,r,peak}$      maximum shaft radial force

$F_{S,r,target}$      target shaft radial force achieved during expansion

$F_{S,z}$      shaft vertical (anchor) force

$F_{Tcone,x}$      tip cone force along x-axis direction

$F_{Tcone,z}$      tip cone vertical force

$F_{Tcyl,top,z}$      vertical force on tip cylinder top surface

$F_{Tcyl,z}$      tip cylinder vertical force

$F_{Tcyl,r}$      radial force measured on tip cylinder

$F_{Tcyl,r,target}$      target radial force on tip cylinder achieved during expansion

$F_{TA}$      tip anchor capacity

$F_{top,z}$      vertical force on the probe top

$g$      gravitational acceleration

$G$      shear modulus

$H_C$      Chamber height

$h_E$      Embeddment height

$h_N$      height of probe neck

$h_S$      height of probe shaft

$h_{TCyl}$      height of probe tip cylinder

| | |
|---|---|
| $m_p$ | probe mass |
| $n_1, n_2, n_3, n_4$ and $n_5$ | scaling factors |
| $N$ | scaling number for gravity |
| $Q_T$ | tip resisting force during penetration |
| $Q_S$ | shaft resisting force during shaft retraction |
| $v_{Tcyl,r}$ | tip cylinder radial expansion velocity |
| $v_{S,r}$ | shaft radial expansion velocity |
| $v_p$ | tip penetration velocity |
| $v_{TO}$ | oscillation velocity of tip point along one single vertical x-z plane |
| $W_{SC}$ | work done by shaft contraction |
| $W_{SE}$ | work done by shaft expansion |
| $W_{SR}$ | work done by shaft retraction |
| $W_{TAC}$ | work done by tip contraction |
| $W_{TAE}$ | work done by tip anchor expansion |
| $W_{TO}$ | work done by tip oscillation |
| $W_{tot}$ | total self-burrowing work |
| $W_{TP}$ | work done by tip penetration |
| $\mu$ | friction coefficient of soil |
| $\mu_p$ | friction coefficient of probe |
| $v$ | Poisson's ratio |
| $\Delta\rho_{shaft}$ | shaft penetration distance in one burrowing cycle |
| $\Delta\rho_{tip}$ | tip penetration distance in one burrowing cycle |
| $\rho_p$ | Probe material density |

# Abbreviations

CPT     cone penetration test
CV      constant velocity
DEM    discrete element method
IP       initial penetration
PRM    particle refinement method
PSD     particle size distribution
SC      shaft contraction
SE      shaft expansion
SR      shaft retraction
TAC     tip anchor contraction
TAE     tip anchor expansion
TP      tip penetration
TPO     tip penetration with oscillation

# 1 Introduction

In recent years, the study of planetary exploration has taken centre stage in the realm of planetary science. This growing interest stems from the potential insights it offers into fundamental questions about our universe. As a result, the focus of planetary exploration has shifted from predominantly surface-oriented studies to more in-depth subterranean investigations [1–3]. Historically, rotation drilling methods, as showcased by the Apollo mission's lunar samples [4], were the standard. However, with the development of advanced rotary percussive drilling devices like the auto-Gopher [5], exploration techniques have evolved. Despite their proven efficacy, these tools present challenges: they can be energy-intensive and often difficult to operate in extraterrestrial conditions [6]. There has been a marked shift towards designing lighter devices [7]; however, the low driving energy available in low gravity environments can lead to relatively shallow penetration depth. Inspiration from nature, specifically from burrowing animals, has led to the design of lightweight and flexible self-burrowing robots. Bioinspired burrowing mechanisms [8] promise the creation of more compact, efficient robots, which appear particularly suited for complex environments.

Bio-burrowing strategies can be broadly classified into wriggling [9], undulating [10], dual-anchoring [11], grabbing-pushing [12], reciprocating [13], granular fluidization [14], and circumnutations and tip lateral expansion [15]. Notably, the dual-anchoring system [16,17], inspired by the burrowing mechanisms of razor clams, is gaining popularity due to its inherent simplicity. This system relies on the clam's upper shell and terminal foot. The shell's expansion acts as the primary anchor, generating the anchorage force necessary for foot penetration. Once embedded, the clam's muscle controls foot expansion to establish the second anchor that allows the rest of the body to move forwards. On average, a clam can attain depths of up to 0.7m, or 5-9 times its body length, using this technique [18].

Experimental and numerical studies have underscored the potential of bioinspired robots for soil analysis [19–21]. For instance, Winter [22] explored burrowing drag reduction mechanisms and devised a robot capable of mimicking such natural processes. Tao [23] created a robot that emulated the clam's burrowing technique, while researchers like Babheri [24] and Zhong [25] indicated that asymmetrical robot designs or kinematics can enhance motions. Discrete element modelling (DEM) simulations [26,27] are particularly insightful in understanding probe-soil interactions, especially to capture multi-scale features. Chen et al., [28] compared one and two-anchor systems, deducing that dual anchors outperform single anchor in terms of anchor efficiency when anchor spacing is smaller than six probe diameters. Additionally, Huang and Tao [29] successfully aligned DEM model outcomes with cavity expansion theoretical predictions regarding energetic cost, which shows dynamic penetration was found to reduce the energetic cost on the penetration of cone and shaft by about 36%, compared with the pure cone penetration strategy.

In our research, we study a dual-anchoring, self-burrowing probe [30] using DEM. Simulations under 1$g$ conditions have shown that the probe could achieve 9.5 cm, one-third of its body length, through repeated 3 burrowing cycles [30]. However, the soil response in extraterrestrial environments differ considerably from those on Earth, mainly due to the varying gravity effects [31–33]. For instance, the stress field and mechanical properties of the soil on the moon will also be different from those on the earth, potentially influencing the interactions with soil exploration probes. Studies like those by Kobayashi [34] and Jiang et al. [31] have highlighted the significant role of gravity in these interactions. They showed that the bearing capacity scales linearly with gravity, comparing under 1/6$g$ and 1$g$. Jiang et al. [31] simulated the penetration resistance using DEM and found a linear relationship between the reciprocal of the gravity level and peak value of the normalized penetration resistance.

Given the high costs, technological demands, and limited testing periods associated with Earth-bound experiments at different gravity conditions, like those in centrifuges and parabolic flights, DEM simulations emerge as an effective alternative to emulate varied gravity conditions. Consequently, two critical research questions arise: Can a self-burrowing probe designed for 1$g$ conditions retain its efficacy in other gravitational environments? Secondly, how do forces, velocities and other variables scale with gravity? In Section 2, the general DEM model, probe and soil chamber are described. Section 3 introduces the working principles of the dual anchor probe, and Section 4 studies the difference in performance under different gravity conditions. Section 5 introduces additional discussions and the paper concludes in Section 6 with the main findings.

## 2   Construction of soil model and probe

In our study, we designed a virtual chamber model within the DEM software PFC3D. This chamber, shaped like a cylinder with dimensions of 0.7m for both diameter and height, contains a soil sample composed of particles replicating the mechanical properties of Fontainebleau sand. Specifically, the sand has a $D_{50}$ of 0.21mm, with particle sizes between 0.1 and 0.4 mm. The maximum and minimum void ratios are 0.9 and 0.51.

To ensure optimized interaction with the probing device while keeping computational efficiency, particles were upscaled using five scaling factors ($n_1$=35, $n_2$=53, $n_3$=79, $n_4$=95, and $n_5$=113) from the central to the peripheral zones, keeping the size distribution of Fontainebleau NE34 sand, as seen in Figure 1a, resulting in a total of 104,320 particles in the model. The thickness in the outmost and innermost zone is 14 cm and the intermediate three zones act as filter layers of thicknesses of 4cm and 5cm [35]. This approach, known as the particle refinement method (PRM) [35,36] as displayed in Figure 1b, reduces computational demands, maintains particle zoning integrity, and ensures a sufficiently large number of contacts between the probe and particles to obtain reliable results. Past research [37] attests to the efficiency of PRM, and Chen and Martinez [38] demonstrated that it does not affect average penetration resistance magnitudes.

The sample creation and particle assemblage followed the process laid out in [30], resulting in an average $D_r$ of 86% and porosity of 0.40 (void ratio of 0.67). This assembly was taken to equilibrium under varied gravity levels: $1/6g$, $1/3g$, $1g$, $5g$, $10g$, and $15g$, resulting in porosities within 4% for all gravities and throughout the sample. The first three represent lunar, Martian, and terrestrial gravity conditions, while the latter three correspond to typical centrifuge testing conditions.

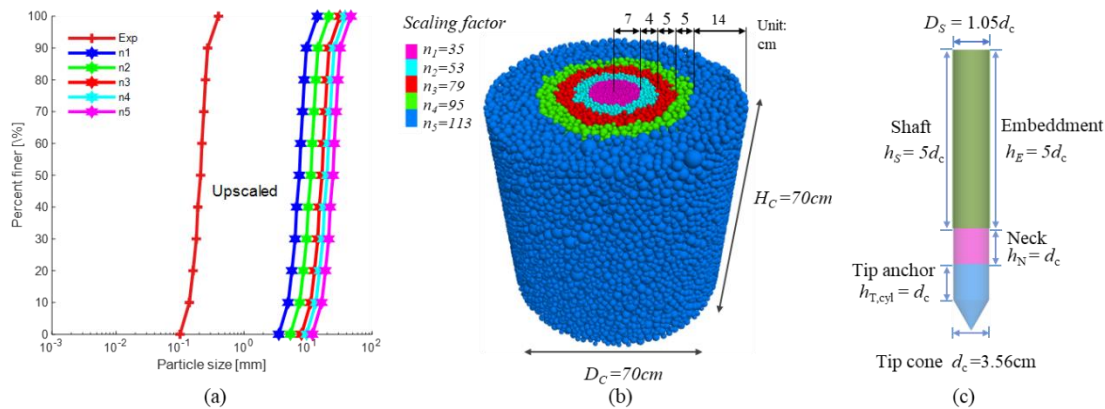

Figure 1 (a) Particle size distribution of Fontainebleau sand, (b) view of soil chamber with multi-scaled particles, and (c) geometry of bio-inspired dual-anchor probe.

We designed a razor clam-inspired dual-anchor probe (Figure 1c) comprising three sections: shaft, neck, and tip. The shaft, starting at 3.74 cm in diameter, can radially expand to generate anchorage forces. The neck, of 3.56 cm in diameter ($d_c$), is designed for vertical movement and connected with the embedment part during penetration, while the 3.56 cm diameter tip includes an expandable base (tip anchor) for anchoring and a 60° apex tip cone, similar to a CPT tests. The segments' lengths of these sections are $5d_c$, $d_c$, and $d_c$, respectively (Figure 1c).

We employed a rough contact model [39], rooted in the standard Hertzian model, calibrated against element tests. This calibrated model, with a shear modulus ($G$) of 32 GPa, Poisson's ratio ($v$) of 0.19, and friction coefficient ($\mu$) of 0.275, restricted particle rotation fully to approximate the effects of particle shape. The simulation's boundaries were frictionless. For an exhaustive calibration process, the reader is referred to [40].

Table 1 DEM contact model parameters

| Element | G (GPa) | v | μ | $S_q$ (μm) | $n_1$ | $n_2$ | ρ |
|---|---|---|---|---|---|---|---|
| F-sand | 32 | 0.19 | 0.275 | 0.6 | 0.05 | 5 | 2.65 g/cm3 |
| Probe | 74 | 0.265 | 0.35 | - | - | - | 8.05 g/cm3 |

Note: $G$, shear modulus; $v$, Poisson's ratio; $\mu$, friction coefficient; $S_q$, surface roughness; $n_1$ and $n_2$, model parameters; $\rho$ sand and probe density.

## 3 Methodology and performance of a stepwise self-burrowing cycle

In this section, we present a brief overview of the self-burrowing methodology proposed in [30] and its efficacy under a 1$g$ conditions. An initial penetration (IP) stage was performed at a constant vertical velocity of 0.4 m/s to attain a depth of 0.34 m [30]. Figure 2 shows two forces during IP: $Q_T$, (sum of the vertical forces from the neck ($F_{N,z}$), tip anchor ($F_{Tcyl,z}$), and tip cone ($F_{Tcone,z}$)) and $Q_S$ (sum of the friction along the shaft ($F_{S,z}$) and the small bearing area that results at the bottom of the shaft ($F_{S,bot,z}$) from the slightly larger diameter of the shaft). Both $Q_T$ and $Q_S$ display an approximately linear increase with depth. Upon reaching the target depth (Point A), the velocity is removed and the probe left to reach force equilibrium under its own weight following Newton's second law, resulting in both $Q_T$ and $Q_S$ swiftly declining to very small values close to zero.

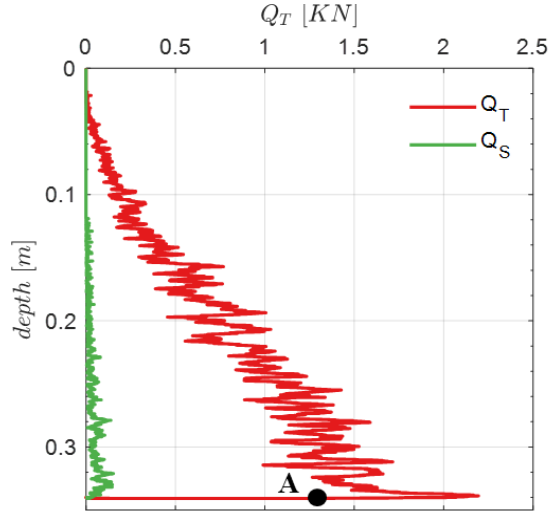

Figure 2 (a) $Q_T$ and (b) $Q_S$ evolution with penetration depth

The self-burrowing cycle consists of six phases governed by force and displacement parameters, as described in Figure 3. These phases are: Shaft Expansion (SE), Tip Penetration with Oscillation (TPO), Tip Anchor Expansion (TAE), Shaft Contraction (SC), Shaft Retraction (SR), and Tip Anchor

Contraction (TAC). Each phase uses specific ending criteria explained later, as shown in Figure 3. Figure 4 details the algorithm's performance via representative forces, where $F_{S,r}$ is the radial force at the shaft and $F_{Tcyl,r}$ is the radial force at the expanded tip. Each of the phases is explained in detail below.

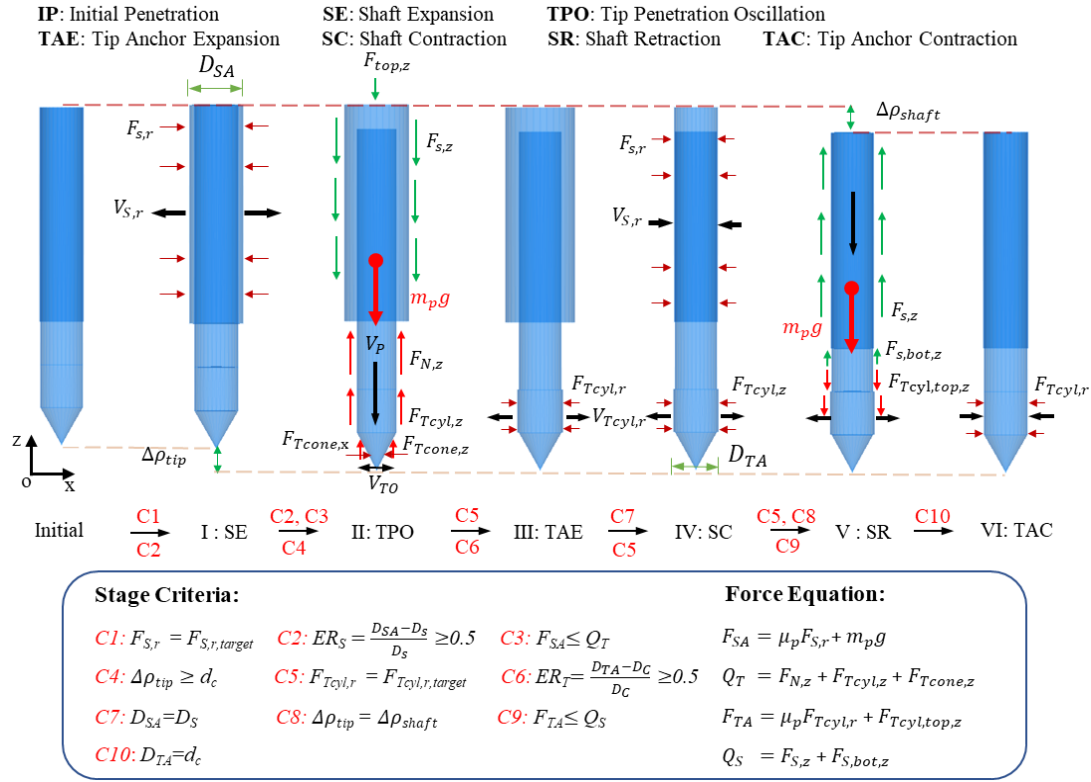

Figure 3 Stepwise self-burrowing cycle and corresponding criteria for each step.

**Step I: Shaft Expansion (SE).** Modelled after the razor clam, the objective is to expand the shaft anchor for subsequent penetration. Using a constant radial velocity of 0.02 m/s, a maximum shaft radial force, $F_{S,r,peak}$, can be calculated. This maximum shaft radial force is the steady-state value (i.e., at which the value is constant for further expansion) experienced by the probe at that velocity. An $F_{S,r,target}$ equal to 80% of of $F_{S,r,peak}$ is used as target shaft radial force, to avoid mobilizing the full passive resistance of the soil. This target force is then divided into 10 increments ($F_{S,r,inc}$) to avoid very large expansion velocities, that are used to gradually expand the shaft. In each increment the load is achieved by adjusting the shaft velocity ($V_{S,r}$) and acceleration which are determined from Newton's second law, concluding when the velocity drops below 0.001 m/s [30]. The expansion ceases either when $F_{S,r}$ matches $F_{S,r,target}$ (*C1*) or when the expansion ratio ($ER_S$) exceeds a functional limit of 50% based on realistic limitations expected for field-testing equipment (*C2*). As shown in Figure 4a, the shaft force $F_{S,r}$ increased gradually until *C1* was triggered, where $ER_S$ is 18%, and hence, much less than the expansion limitation defined in *C2*.

**Step II: Tip Penetration with Oscillation (TPO).** During tip penetration, the shaft continues expanding to maintain the required $F_{S,r,target}$ set in Step I, and oscillation of the tip cone is adopted to reduce $Q_T$. In one oscillation cycle, the cone tip point firstly moves horizontally to the far left in the x-z plane with $v_{TO}$ of 0.8 m/s during 0.025 s. Then the point moves in the opposite direction to the far right during 0.05s at the same velocity. Finally, the velocity is reversed again and the tip point returns to the original middle position after 0.025s. The purpose of the oscillations is to reduce the penetration resistance, as shown in [30]. The neck and tip (anchor and cone – see Fig. 1 and 3) move downwards a distance of $\Delta\rho_{tip}$ with a velocity of 0.05 m/s ($v_p$). Three criteria are checked: *C3*, when the available reaction to push forward ($F_{SA}$) becomes smaller than the total tip penetration resistance $Q_T$; *C4*, when the penetration distance exceeds one cone diameter; and *C2*, when the shaft expansion ratio exceeds 50%. *C4* is also a functional limit imposed following the criteria of *C2* in the previous stage. In this simulation, the SPO stage is terminated when *C2* is triggered (see Figure 4c). Due to the combined horizontal and vertical movement of the tip, $Q_T$ oscillates in magnitude. The shaft force $F_{S,r,target}$ was kept constant until *C2* was triggered.

**Step III: Tip Anchor Expansion (TAE).** Similar to shaft expansion, the tip anchor expands to a final diameter of $D_{TA}$ to form the second anchor. The same incremental loading algorithm introduced in Step I is again used to reach the tip expansion target force $F_{Tcyl,r,target}$ which is determined based on the maximum $Q_s$ during the initial penetration process and $Q_T$ at the end of TPO stage (($Q_s + Q_T)/\mu$). In this step, there are two termination criteria: *C5*, when the tip anchor radial force ($F_{Tcyl,r}$) reaches $F_{Tcyl,r,target}$; and *C6*, which is similar to *C2*, when the tip anchor expansion ratio ($ER_T$) reaches 50%. After the execution of this stage, $F_{Tcyl,r}$ increased until *C5* was triggered, while $ER_T$ increased by a relatively small magnitude of 4% (Figure 4). It can be seen in the same figure that during tip expansion, $F_{S,r}$ significantly decreased due to the interactions between the shaft and tip.

**Step IV: Shaft Contraction (SC).** The shaft is contracted to its original diameter at a constant rate of 0.1 m/s. During this step, the tip anchor continues to expand to maintain $F_{Tcyl,r,target}$ (*C5*). When $D_{SA}$ returns to its original size (*C7*), contraction is stopped, and the value of $ER_S$ decreases to zero. Due to the interactions between the different probe segments, $F_{Tcyl,r}$ first dropped rapidly, and then increased to $F_{Tcyl,r,target}$ after only 7% of tip anchor expansion.

**Step V: Shaft Retraction (SR).** During the shaft retraction process, the shaft is dragged downwards at a constant velocity of 0.1 m/s until the shaft penetration distance $\Delta\rho_{shaft}$ is equal to $\Delta\rho_{tip}$ (*C8*). In this stage, the $F_{Tcyl,r}$ is still kept constant (*C5*) and we check that the tip anchor capacity ($F_{TA}$) is greater than the shaft retraction resistance $Q_S$ (*C9*). At this point, the probe length returns to its original value.

**Step VI: Tip Anchor Contraction (TBC).** This stage ends when the tip contracts back to its original size at a constant velocity of 0.1 m/s, *C10*. The three representative forces drop to values near zero. By the end of the self-burrowing cycle, the probe penetrated 2 cm successfully without any external assistance.

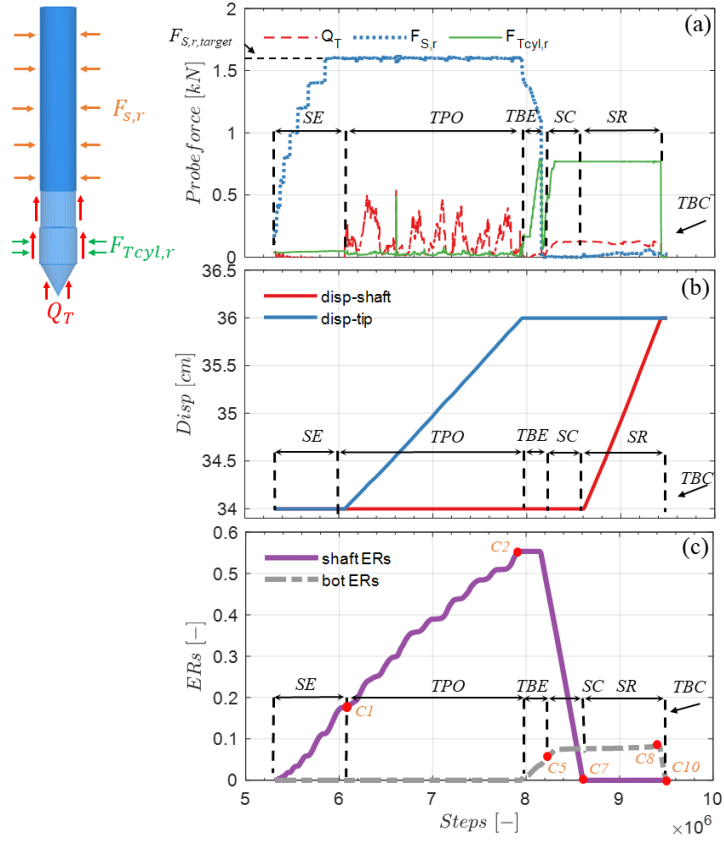

Figure 4 Evaluation of (a) characteristic forces, (b) shaft and tip displacement and (c) expansion ratio recorded in a complete cycle in 1$g$ condition.

## 4  Gravity effects on critical self-burrowing steps

In this section, we investigate the effects of gravity level on critical self-burrowing steps, including Initial Penetration (IP), Shaft Expansion (SE), Tip Penetration with Oscillation (TPO), and Tip Anchor Expansion (TAE). It is observed that the evolution of the forces relevant during the SC, SR, and TAC stages ($Q_T$, $F_{S,r}$, and $F_{Tcyl,r}$) are not affected by gravity. Therefore, they are not presented here for the sake of brevity. In these simulations, six gravities used, namely 1/6$g$, 1/3$g$, 1$g$, 5$g$, 10$g$ and 15$g$, resulting in gravitational scaling factors $N$ of 1/6, 1/3, 1, 5, 10 and 15, where 1$g$ represents Earth's gravity.

For every model, the length dimensions ($L$) and density ($\rho$) are maintained equal, and hence the scaling factor is 1.0, whilst gravity is scaled by $N$, as previously mentioned. Hence, based on a simple dimensional analysis we anticipate that the scaling factors for other parameters should be as presented in Table 2. We only show variables that are relevant to this paper. The scaling for energy ($E$) is estimated only for potential and kinetic energy.

*Table 2. Scaling factors assuming all linear and superposition applies.*

| Variable | Units | Scaling factor, $N = X_{new\text{-}g} / X_{1\text{-}g}$ |
|:---:|:---:|:---:|
| L | m | 1.0 |
| $\rho$ | kg/m$^3$ | 1.0 |
| g | m/s$^2$ | N |
| F | N | N |
| v | m/s | $N^{1/2}$ |
| E | J | N |

## 4.1 Initial penetration

Figure 5 and

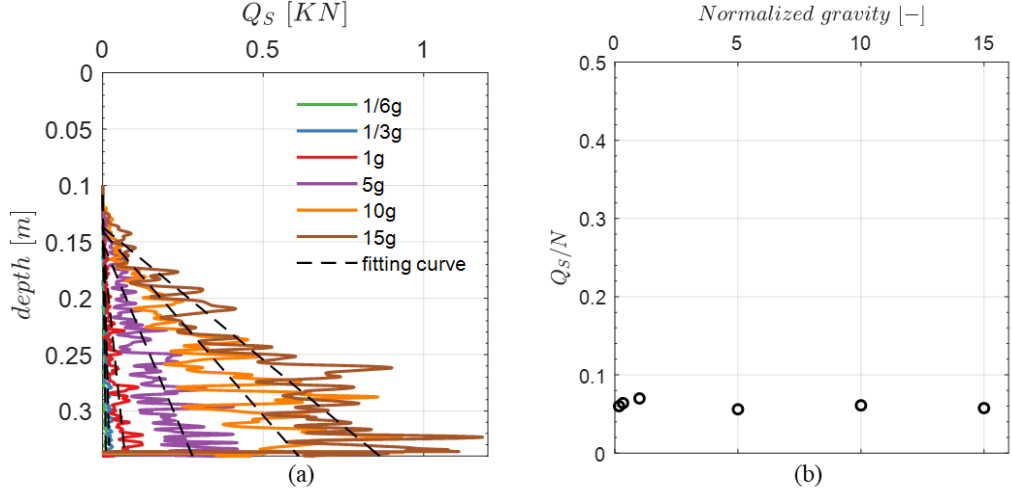

Figure 6 present the evolution of $Q_T$ and $Q_S$ against penetration depth, respectively. $Q_T$ increases nearly linearly with penetration depth under different gravity conditions. $Q_T$ increases with gravity (Figure 5a), in agreement with the general observations reported in Jiang et al [31] and findings from static penetrating centrifuge model test [41]. $Q_S$ also increases with increasing depth and gravity (see Figure 6b). The maximum value of $Q_S$ is also important because it is used for determining the value of $F_{Tcyl,r,target}$ in the tip anchor expansion stage.

The proposed scaling factor $N$ for forces are confirmed in Figures 5b and 6b, which shows normalized forces at the end of penetration. The results are reasonable and fall within 8% for this stage.

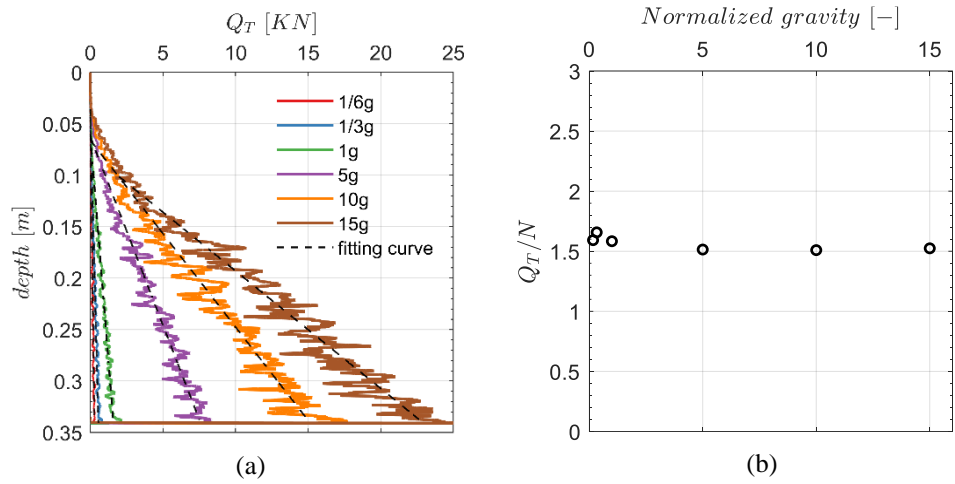

Figure 5 $Q_T$ evolution with penetration depth in all gravity (a) and scaled $Q_T$ (b).

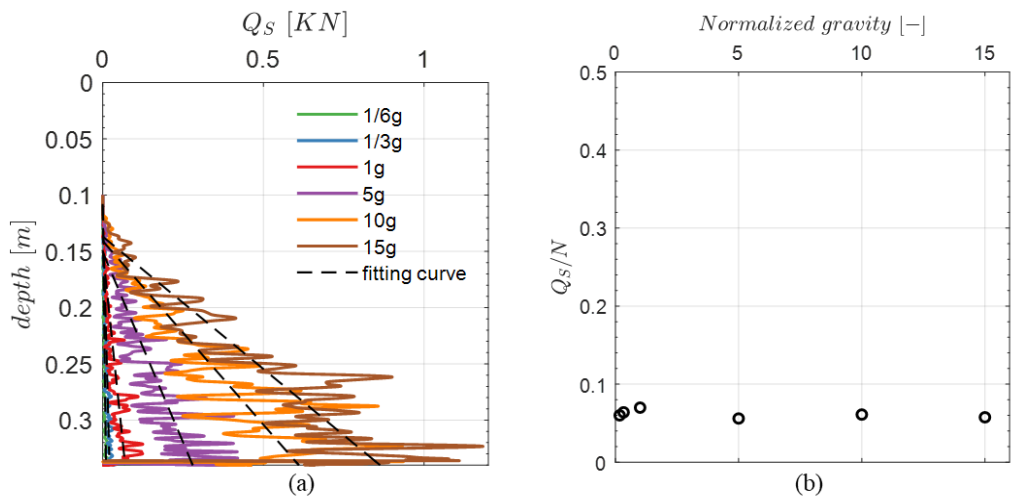

Figure 6 $Q_S$ evolution with penetration depth in all gravity (a) scaled $Q_S$ (b).

## 4.2 Shaft expansion

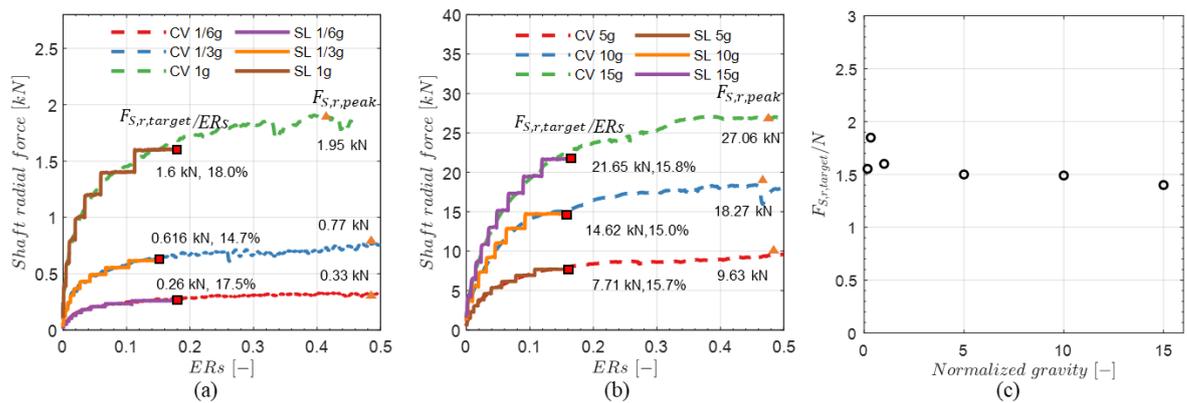

Figure 7 shows the shaft radial force against expansion ratio under low and high gravity levels. The SE stages were terminated after triggering criterion *C1* for all gravity levels. The dashed lines shown in the figure were first obtained through shaft expansion simulations at constant velocity to determine the peak force $F_{S,r,peak}$ to subsequently determine the target force (i.e., $F_{S,r,target} = 0.8F_{S,r,peak}$). The solid lines represent the incremental staged loading that was applied to reach the target forces. In general, with the increase of gravity, $F_{S,r}$ increases, leading to a linear increase of the target forces. The maximum target forces are 0.264 kN, 0.616 kN, 1.6 kN, 7.5 kN, 14.9 kN, 21 kN for increasing gravities, respectively.

The final expansion ratio ranged from 0.147 to 0.18 which is well below the imposed functional limit of 0.5. The small differences in $ER_S$ indicates that the expansion magnitude that is required to generate an anchorage force equal or greater than the penetration force is independent of gravity. Thus, although the soil stress varies linearly with gravity (by the scaling factor *N*), the performance of shaft anchor in terms of the required expansion is not significantly affected. In fact, the $EM_S$ values being smaller than 0.2 allow for significant additional expansion during the next TPO stage.

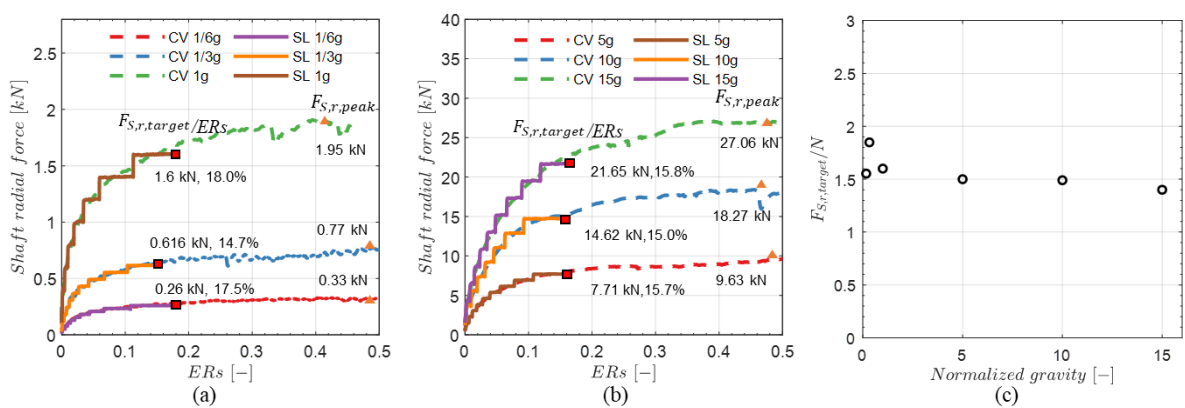

Figure 7 Shaft radial force against expansion ratio in (a) low, (b) high gravity levels and (c) scaled target force

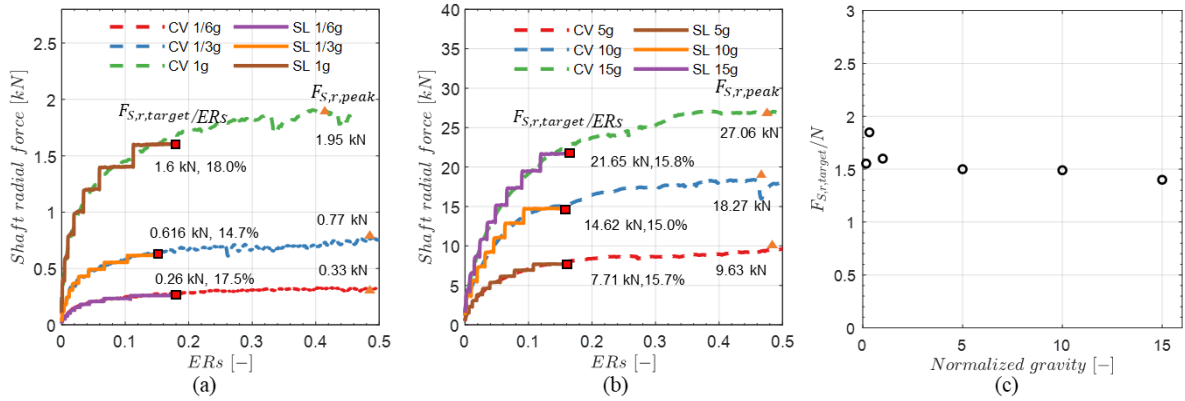

Figure 7c again confirms the scaling factor. The differences between low and high gravity is attributed to the fact that for low gravities the soil has fully yielded forming a full failure wedge, whereas for the higher gravity this is not the case (see Fig. 8d).

The typical shaft expansion velocity and its average velocity against shaft expansion ratio is plotted in Figure 8(a) for the 1/6$g$, as an example. In each incremental stage, the velocity is adjusted and increases quickly to a peak to maintain the shaft target force, and then drops for stabilization. Figure 8(b) shows the average radial velocity for all gravity conditions against for a corresponding normalized gravity of 1$g$. The velocities increase with gravity, with values of 0.008m/s, 0.01m/s, 0.019m/s, 0.027m/s, 0.038m/s, 0.037m/s for the 1/6$g$, 1/3$g$, 1$g$, 5$g$, 10$g$, and 15$g$ cases, respectively.

Fig. 8c shows the applied scaling factor from Table 2. It shows that the datapoint clusters in two groups; one for low gravity conditions and one for high gravity. The particle displacement field shown in Fig. 8d explain this. It shows the final particle displacement field at the end of SE step along the center x-z cross-section for 1/6$g$ and 5$g$ conditions. A pronounced shallow conical failure zone for 1/6g was formed so that at the full-strength mobilization developed for low gravities, larger velocities of the shaft are necessary to maintain a constant load and the scaling is not expected to work. And small particles displacement for high gravity shows low expansion velocity is enough to constant force.

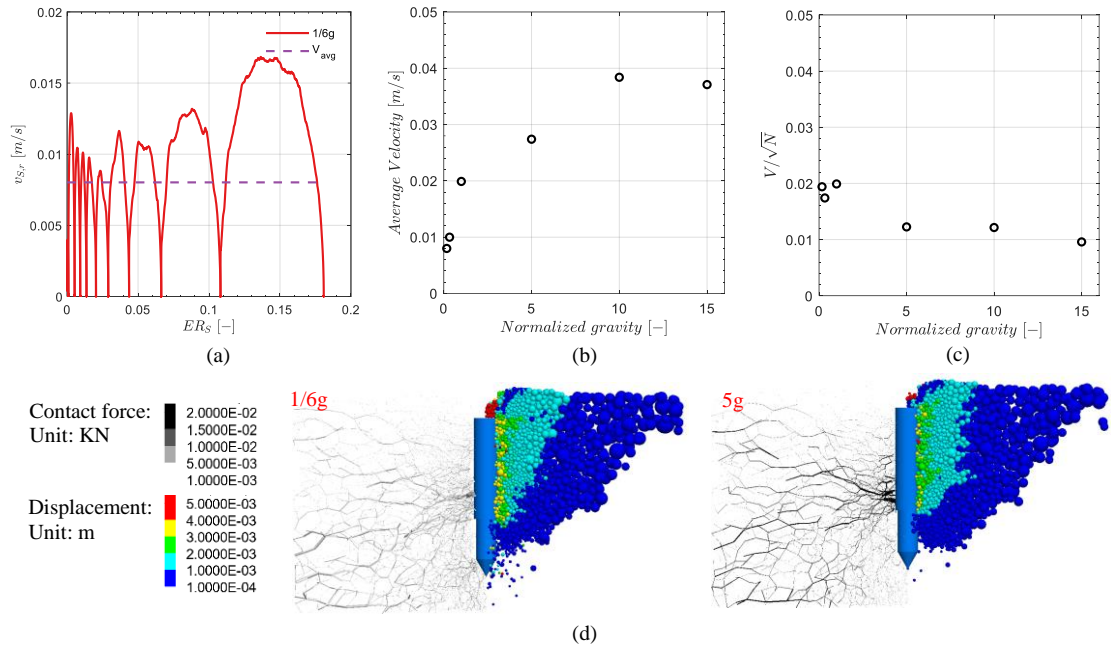

Figure 8 (a) Radial expansion velocity against expansion ratio in the case of 1/6g, (b) the average velocity in the SE stage against normalized gravity, (c) scaled average velocity, (d) Contact force chain (left part of each subplot) and ball displacement (right part of each subplot) in the SE with 1/6g and 5g condition

### 4.3 Tip penetration with oscillation

The aim of developing self-burrowing robots is to penetrate to shallow or deep soils without assistance of an external force. Therefore, penetration depth is a key factor to evaluate gravity effects on the self-burrowing performance. Figure 9 shows the evaluation of shaft expansion ratio against depth under increasing gravity. It is noted that the probe achieves a greater advancement of 4cm under high gravity than that of 2cm under low gravity. The penetration vertical velocity is 0.05m/s in all simulations and hence, the corresponding time is 0.4s in low gravity and 0.8s in high gravity.

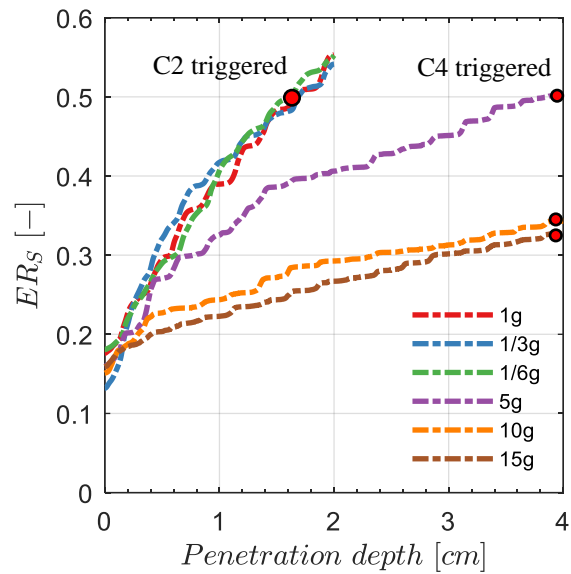

Figure 9 Shaft radial expansion ratio against penetration depth.

At the end of the TPO stage, the final $ER_S$ value is an important factor for the final penetration depth. The ratios at the beginning of TPO are similar, but the final $EM_S$ are very different between low and high gravity conditions. The final shaft expansion ratios under low gravity reached the limit of 50%, thus triggering the $C2$ termination criterion. In contrast, the $EM_S$ values at the end of penetration for the $5g$, $10g$, and $15g$ cases are 50%, 34%, and 32% respectively. The depth of 4cm in this cases means the displacement criteria ($C4$) was triggered first.

The typical shaft expansion velocity and its average velocity for the $1/6g$ case are plotted against depth in Figure 10 (a). Figure 10 (b) shows the average velocity during the TPO stage against all gravity conditions. The average shaft expansion velocity is 0.0176 m/s, 0.0195m/s and 0.0181m/s for the $1/6g$, $1/3g$, $1g$ simulations, respectively, while for $5g$, $10g$ and $15g$ the velocity has lower values of 0.0084m/s, 0.0046m/s and 0.004m/s, reflecting the greater rate of change at low gravity values.

The tip oscillation and penetration can weaken the force chain between the soil particles and the shaft anchor, as shown previously by [28]. In low gravity environments, the shaft expands more rapidly to maintain the target force. However, there is relatively a smaller influence on the force chains under high gravity [42]. These results indicate that at gravities of $1g$ and below, the limiting factor for penetration is the anchorage capacity, which triggers $C2$. For larger gravity values such as $10g$ and $15g$, the limiting criteria is the vertical penetration, whilst there remains a significant capacity to further mobilize additional shaft resistance. Using the analogy of higher gravity having similar effects as a greater density, this confirms the results of Chen et al [43].

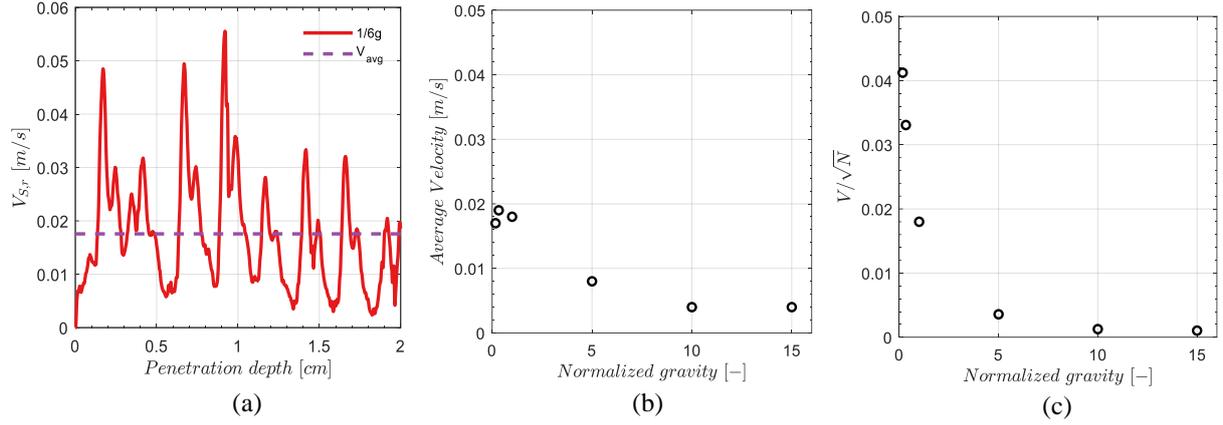

Figure 10 Evaluation of shaft radial velocity against depth in the 1/6g. (b) the evolution of shaft expansion velocity against normalized gravity in TPO stage. (c) scaled velocity

The scaling factor for shaft velocity (Fig. 10c) is clearly not so well fulfilled because the shaft velocity is a non-local effect caused by the interactions at the tip and therefore, the simple scaling is not expected to work. This is phenomenologically confirmed when we plot the scaled velocity and a power-law scaling emerges, which is typical of non-local effects in granular materials [44].

### 4.4 Tip anchor expansion

Figure 11 shows the final tip anchor expansion ratio and the appropriateness of the scaling of the corresponding radial force. For all gravity levels, the value $ER_T$ is well below the final limit of 0.5, (Figure 11a). The corresponding tip anchor target force ($F_{Tcyl,r,target}$) becomes larger with the increasing gravity. As shown in Figure 11b the proposed scaling factor collapses the data relatively well. However, the differences between low and high gravity are attributed to the fact that the penetration depth is greater for the high gravity models (i.e., 2 vs 4cm).

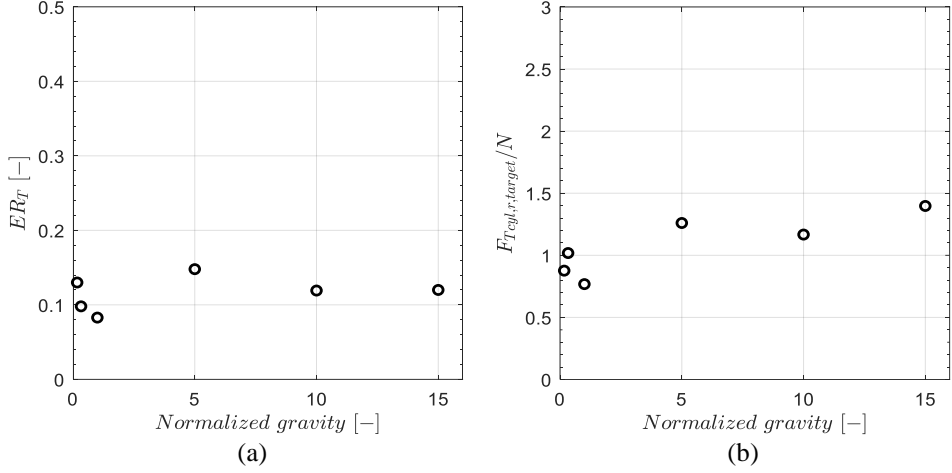

Figure 11 (a) Tip anchor expansion ratio and (b) scaled bottom target force.

# 5 Work done during all self-burrowing steps

The total amount of energy used during one self-burrowing cycle is crucial for the design of a robotic device. The total work $W_{tot}$ is calculated using Equations (2-6), and consist of the work done during seven motions: shaft expansion ($W_{SE}$), tip cone penetration ($W_{TP}$), tip oscillation ($W_{TO}$), tip anchor expansion ($W_{TAE}$), shaft contraction ($W_{SC}$), shaft retraction ($W_{SR}$) and tip anchor contraction ($W_{BTAC}$). The energy of shaft motion and tip anchor motion can be easily represented with the Equation 2 and 5 in the SE, TPO, TAE and TAC stages, because the calculation of expansion and contraction is the same.

$$W_{tot} = W_{SE} + W_{TP} + W_{TO} + W_{TAE} + W_{SC} + W_{SR} + W_{TAC} \quad (1)$$

$$W_{SE} = W_{SC} = \int |F_{S,r}(t) v_{S,r}(t)| dt \quad (2)$$

$$W_{TP} = \int |Q_T(t) v_p(t)| dt \quad (3)$$

$$W_{TO} = \int |F_{Tcone,x}(t) v_{TO}(t)| dt \quad (4)$$

$$W_{TAE} = W_{TAC} = \int |F_{Tcyl,r}(t) v_{Tcyl,r}(t)| dt \quad (5)$$

$$W_{SR} = \int (|F_{S,z}(t) v_p(t)| + |F_{S,bot,z}(t) v_p(t)|) dt \quad (6)$$

where $v_{S,r}$ is the radial velocity of the shaft, and $v_p$ is the vertical velocity of the tip, $F_{Tcone,x}$ and $v_{TO}$ are the radial force along the x-axis direction and the tip oscillation velocity, $v_{Tcyl,r}$ is the radial velocity of the bottom shaft.

shows the total work ($W_{TOT}$) done by the complete self-burrowing stages, showing that the two largest work components are the tip oscillation ($W_{TO}$) and shaft expansion ($W_{SE}$). In the low gravity conditions, the tip oscillation does between 63% and 67% of the work while shaft expansion does about between 25% and 29%. In contrast, in high gravity conditions the proportion of $W_{TO}$ increases a range between 75% and 79.5%, and the proportion of $W_{SE}$ decreases to the values between 8.7% to 15%. In general, tip oscillation and shaft expansion consumed over 90% of the total energy.

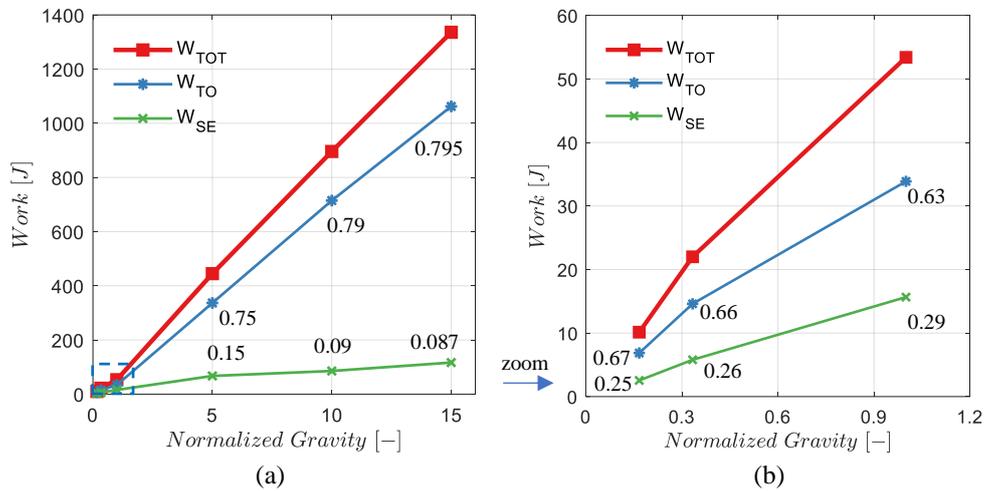

Figure 12 Gravity effects on total work and the main work consumption in TO and SE. The labels in the figure indicate the proportion of work done relative to the total work done.

shows that the proposed scaling $N$ works well for all $W_{TOT}$, $W_{TO}$, and $W_{SE}$, although it forms two groups, one for low gravities below 1$g$, and another for values larger than that. We justify this due to the different energy mixes shown above. Whilst $W_{SE}$ scales well with $N$, $W_{TO}$ presents more differences between the different groups, mainly due to the larger penetration in the higher gravity simulations. In fact, when we plot the scaled energy for high gravity at 2cm, the scaling agrees with the low gravity values much better.

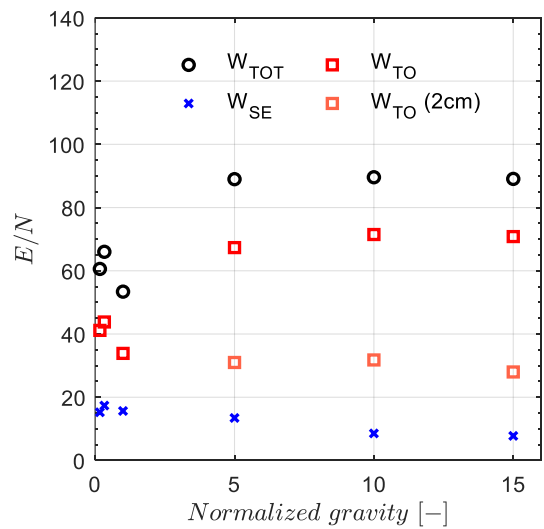

Figure 13 the scaling number for the work consumption in tip oscillation ($W_{TO}$) and shaft expansion ($W_{SE}$).

Figure 14 shows that the total work grows with the increasing gravity. The largest work for one complete self-burrowing cycle is 1341.26 J under 15$g$, and the lowest work is just 10.14 J under 1/6$g$. The penetration efficiency, plotted in Figure 14a quantifies the gravity effect on performance of self-burrowing step in terms of cm of penetration per J, as done by Chen et al. [42]. It shows that the efficiency decreases with increasing gravity. For energy efficiency, the proposed relationship in Table 2 also unifies the data relatively well, as shown in Figure 14b.

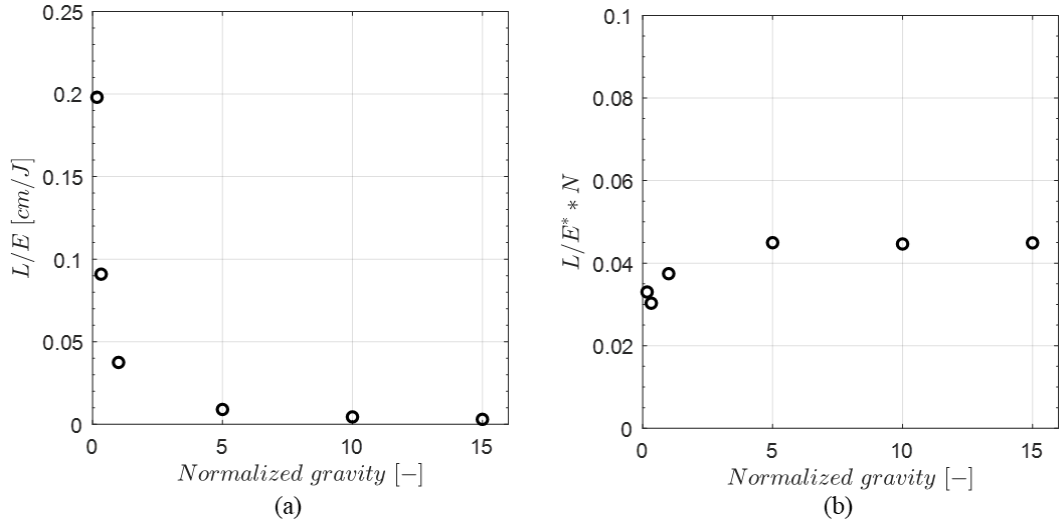

Figure 14 (a) Total energy and energy efficiency against increasing gravity, (b) Scaled energy efficiency

# 6  Conclusion

This study explored the feasibility of the operation of a bio-inspired self-burrowing probe in environments with various gravity levels, namely, 1/6$g$, 1/3$g$, 1$g$, 5$g$, 10$g$ and 15$g$. To ensure the successive execution of the self-burrowing process, force-control algorithms and a tip oscillation strategy are adopted to maintain the reaction force and reduce the penetration resistance. The self-burrowing process consists of six individual steps: shaft expansion, tip penetration with oscillation, tip expansion, shaft contraction, shaft retraction, and tip contraction.

The results show that the probe is capable of self-burrowing under different gravity conditions. However, it can reach greater depths at higher gravity values. The reason is the more efficient lateral confinement provided by the soil which allows penetrating without reaching the imposed functional expansion magnitude limit of 50% shaft expansion and the full wedge failure. This proves additional capacity for further penetration.

Additionally, we observe that all first order effects can be scaled using dimensional analysis with gravity as the scaling parameter. This includes forces, velocities, expansion ratios and energy. Small differences in the scaling are observed and are attributed to the fact that low gravity samples mobilise the full strength of the material, as opposed to high gravities. This results in two clusters; one for gravities equal and greater than $5g$ and another for values equal and smaller than $1g$. However, the normalized values only differ by 22% for forces, 40% for velocities and 33% for total energy.

Tip oscillation consumes more than 60% of the total energy of the probe. The total energy and the penetration depth energy efficiency present opposite trends, i.e., the greatest efficiency is achieved for the lowest gravity, $1/6g$, while the largest energy consumption occurs for $15g$ as expected. In terms of future applicability, the study confirms that the probe could be used in gravitational environments representative of outer-space bodies such as the Mars and the Moon with lower gravity on or higher gravity applications such as centrifuge testing. Additionally, the proven scaling factor can be used to design the probe for different conditions.

# 7  Acknowledgements

The first author has received support from a CSC grant, which is greatly acknowledged. The second author thanks the financial support of the Theodore von Kármán Fellowship - outgoings 2023 (GSO082) from RWTH Aachen University. This material is based upon work supported in part by the Engineering Research Center Program of the National Science Foundation under National Science Foundation (NSF) Cooperative Agreement No. EEC–1449501. The third and fourth authors were also supported by the NSF under award No. 1942369. Any opinions, findings, and conclusions or recommendations expressed in this material are those of the author(s) and do not necessarily reflect those of the National Science Foundation.